\documentclass[12pt]{article}

\usepackage{setspace}
\usepackage{amsmath, amssymb, amsthm}

\def\K{{\mathcal K}}

\def\be{\begin{equation}}
\def\ee{\end{equation}}
\def\lp{\ell_P}
\def\R{{\mathcal R}}
\def\R{{\mathcal R}}

\def\K{{\mathcal K}}
\def\be{\begin{equation}}
\def\ee{\end{equation}}
\def\lp{\ell_P}

\def\beq{\begin{eqnarray}}\def\eeq{\end{eqnarray}}
\begin{document}
\title{{\bf On generalized gravitational entropy, \\ squashed cones and holography} }
\author{ Arpan Bhattacharyya$^{1}$, Menika Sharma$^{1,2}$ and  Aninda Sinha$^{1}$\\
\it $^{1}$ Centre for High Energy Physics,
\it Indian Institute of Science,\\ \it C.V. Raman Avenue, Bangalore 560012, India. \\
$^{2}$\it Harish-Chandra Research Institute\\
 \it  Chhatnag Road, Jhusi,
Allahabad 211019, India
\date{}}

\maketitle
\vskip 1cm
\begin{abstract}{\small We consider generalized gravitational entropy in various higher derivative theories of gravity dual to four dimensional CFTs using the recently proposed regularization of squashed cones. We derive the universal terms in the entanglement entropy for  spherical and cylindrical surfaces. This is achieved by constructing the Fefferman-Graham expansion for the leading order metrics for the bulk geometry and evaluating the generalized gravitational entropy. We further show that the Wald entropy evaluated in the bulk geometry  constructed for the regularized squashed cones leads to the correct universal parts of the entanglement entropy for both spherical and cylindrical entangling surfaces. We comment on the relation with the Iyer-Wald formula for dynamical horizons relating entropy to a Noether charge. Finally we show how to derive the  entangling surface equation in Gauss-Bonnet holography.  }
\end{abstract}
\tableofcontents
\onehalfspace

\section{Introduction}
Recently Lewkowycz and Maldacena (LM) \cite{maldacena} have proposed a derivation of the Ryu-Takayangi (RT) prescription \cite{ryu} for computing entanglement entropy (EE) \cite{calabrese} in holography \cite{rev}. A generalization of black hole entropy is proposed in the context where there is no U(1) symmetry in the bulk. In the Euclidean theory, although there is no U(1) symmetry, one imposes a periodicity condition of $2\pi n$ with $n$ being an integer on the Euclidean time direction at the boundary. This time direction shrinks to zero at the boundary. By suitably choosing boundary conditions on the fields, LM propose to identify the on-shell Euclidean action with a generalized gravitational entropy.

In calculations of entanglement entropy in quantum field theories, one frequently uses the replica trick which entails introducing a conical singularity in the theory \footnote{The only example where a derivation of EE exists without using the replica trick is for the spherical entangling surface \cite{chm, myersme} although in \cite{smolkin} it has been explained how this procedure is connected with the replica trick. A proposal has been made in \cite{AB, area, ab} for the equation for the entangling surface which does not depend on the replica trick.}. An earlier attempt to prove the Ryu-Takayanagi formula was made by Fursaev \cite{fursaevd}. In recent times, in the context of AdS$_3$/CFT$_2$ there have been further developments in \cite{bum1,bum2} towards a proof. In the context of holography, this corresponds to taking the $n\rightarrow 1$ limit. In this case, LM suggest that the time direction shrinks to zero on a special surface. The equation for this surface is derived in Einstein gravity by showing that there is no singularity in the bulk equations of motion. This surface has vanishing trace of the extrinsic curvature and corresponds to a minimal surface--which is precisely what comes from minimizing the Ryu-Takayanagi area functional.

The identification of the entanglement entropy with the generalized gravitational entropy opens the avenue for systematically generalizing holographic entanglement entropy for more general bulk theories of gravity other than Einstein gravity. This understanding is crucial in order to understand systematics of how finite coupling effects in the field theory  modify entanglement entropy. There are two kinds of corrections: a) those which arise from ``classical" and local higher derivative corrections to the bulk theory and b) those which arise from ``quantum" or loop corrections to the effective action which would include non-local effects \cite{quantum}. In this paper we will focus on the former. 

In \cite{ab} (see also \cite{Chen}) we extended the LM method for deriving the entangling surface equation to four derivative gravity. We found that in order for the method to be applicable we needed the extrinsic curvature to be small and in this regime, the surface equations for Gauss-Bonnet gravity coincided with that in the literature. In particular it coincided with what arises from the Jacobson-Myers (JM) entropy functional \cite{jm} which differs from the Wald entropy functional \cite{wald} in terms quadratic in the extrinsic curvature \cite{higher, higher2}. For a more general four derivative action, we could not find a suitable entropy functional.

In a parallel development, Fursaev et al \cite{solo} proposed an extension of the regularization of conical singularities \cite{fursaev} to surfaces having extrinsic curvature--which they call squashed cones. In that paper, they proposed an entropy functional which is supposed to be applicable for an arbitrary four derivative theory. As a check, their functional coincides with the JM entropy functional for the Gauss-Bonnet case. In this paper we will show that indeed their entropy functional gives the expected \cite{solod} universal terms for spherical and cylindrical entangling surfaces in arbitrary four derivative theories. This motivates us into looking at the following questions:

\begin{itemize}
\item Does the new regularization give the correct universal terms when calculated using the generalized gravitational entropy? In order to answer this question we will need to compute the bulk metric using the Fefferman-Graham metric.

\item Is there a relation between entanglement entropy and Wald entropy?

\item Can this lead to the expected equation for the entangling surface in Gauss-Bonnet gravity without needing to resort to a weak extrinsic curvature limit? What light does this shed on the LM method?
\end{itemize}

Let us summarize our findings to the questions above. Indeed we find that the new regularization of Fursaev et al leads to the expected universal terms in the EE for spherical and cylindrical entangling surfaces. In  order to do the computation using the generalized gravitational entropy approach, we need to start with the boundary metric in the form given in \cite{solo}. Then we compute the Fefferman-Graham expansion to leading order. In effect we are computing (upto an order) the bulk metric with a dual which corresponds to field theory living in the entangling region. The resulting bulk metric will be singular. However, in the language of \cite{maldacena} these singularities are mild. In particular in the $n\rightarrow 1$ limit, they will not show up in the on-shell action. Furthermore for Gauss-Bonnet gravity as we will explain, this new regularization indeed leads to the surface equation being the same as that coming from the JM entropy functional. We will explain that a modification to the order of limits needs to be done to the use of the  LM method in \cite{ab} for deriving the surface equation. 

We also address the connection of EE with Wald entropy \cite{wald}. As is by now well known, in an arbitrary theory of gravity, taking the Wald entropy functional in AdS space will give rise to the wrong universal terms in EE \cite{higher, higher2}. In Gauss-Bonnet gravity, the correct entropy functional is the JM one. This was obtained using a Hamiltonian approach. Unfortunately, this makes it really hard to guess a suitable entropy functional for an arbitrary theory of gravity.  The approach of \cite{solo} may be a way around this problem. However, for an arbitrary higher derivative theory of gravity, it entails first working out the entropy functional and then working with it--currently, this has been possible only at the four derivative level. The advantage of the Wald formula in the context of black holes was that it was applicable for any theory of gravity with arbitrary higher derivative corrections. Unfortunately, in the Noether charge method which leads to the Wald entropy, there are ambiguities which can only be resolved for bifurcate horizons \cite{Kang, Iyer}. Iyer and Wald \cite{Iyer} had proposed a prescription that generalizes the Wald entropy to dynamical horizons which are not bifurcate. The prescription is to construct a new bulk spacetime in which the dynamical horizon becomes a bifurcate Killing horizon for which the extrinsic curvatures vanish. Then one computes the usual Wald entropy in this spacetime. The resulting entropy functional for Lovelock theory coincides with JM. The key feature that made this possible was the construction of the new spacetime in which the extrinsic curvatures of the original surface vanish. We will find that the Fefferman-Graham metric, for the cases where the boundary metrics are given by the regularized metrics proposed in \cite{solo}, has some similarities with the Iyer-Wald construction. In particular, in the order of limits proposed in \cite{solo}, the extrinsic curvatures for the entangling surface vanish. This leads to the expectation that the Wald entropy in the bulk spacetime will lead to the correct universal terms. We show that this expectation is indeed true, provided we choose a particular regularization\footnote{Various consistency checks are performed in section 4.}. This regularization will turn out to be surface dependent but theory independent. At the onset, we should clarify that there is no contradiction with the statement above that the Wald entropy functional in AdS space does not lead to the correct universal terms in EE. In the calculation we do, the Wald entropy is computed in the Fefferman-Graham metric which is that of a spacetime which is only asymptotically AdS and with the boundary that of a regularized squashed cone. 

The paper is organized as follows. In section 2, we show that the entropy functional in \cite{solo} leads to the expected universal terms for cylindrical and spherical entangling surface. In section 3, we compute the generalized gravitational entropy in various higher derivative gravity theories. In section 4, we show that the Wald entropy evaluated in the bulk constructed using the Fefferman-Graham expansion leads to the expected universal terms for both the spherical and cylindrical entangling surfaces. In section 5, we revisit the derivation of the entangling surface following \cite{maldacena} in light of the regularization proposed in \cite{solo}. We conclude in section 6.  We have used the same curvature convention as in \cite{wald} throughout our paper.


\section{Entropy functional for general $R^{2}$ theory}
 We will first consider the recently proposed entropy functional for a general four derivative gravity theory  \cite{solo} for a four dimensional CFT.
We will write the bulk AdS metric as
\be \label{metric}
ds^{2}=\frac{\tilde L^{2}}{ z^{2}}(dz^{2}+d\tau^{2}+h_{ij}dx^{i}dx^{j})
\ee
where, $\tilde L$ is  the AdS radius and $h_{ij}$ is a three dimensional metric given below. We will use Greek letters for the bulk indices and Latin letters for the three  dimensional indices.  For the calculation of EE  for a spherical` entangling surface we write the boundary $h_{ij}$ in spherical polar coordinates as,
\be
^{sphere}h_{ij}dx^{i}dx^{j}= d\rho^{2}+\rho^{2}d\Omega_{2}^{2}\,,
\ee 
where $d\Omega_{2}^2=d\theta^{2}+\sin^2 \theta d\phi^{2}$ is the metric of a unit two-sphere and
$\theta \in [0,\pi]$ and $\phi\in [0, 2\pi].$\\
For a cylindrical entangling surface,
\be
^{cylinder}h_{ij}dx^{i}dx^{j}=du^{2}+d\rho^{2}+\rho^{2}d\phi^{2}\,.
\ee
$u$ is the coordinate along the direction of the length of the cylinder. For a cylinder of length $H$, $u\in [0,H].$ 
Here $\tilde L=\frac{L}{\sqrt{f_{\infty}}}.$

The lagrangian  for a general $R^{2}$ theory,
\be \label{R}
S =-\frac{1}{2\lp^{3}}\int d^{5} x\sqrt{g}\big[R+\frac{12}{L^{2}}+\frac{L^{2}}{2}(\lambda_{1}R_{\alpha\beta\mu\nu}R^{\alpha\beta\mu\nu}+\lambda_{2}R_{\alpha\beta}R^{\alpha\beta}+\lambda_{3}R^{2})\big]\,.
\ee
In this case, $f_\infty$ satisfies $1-f_\infty+\frac{1}{3}f_\infty^2 (\lambda_1+2\lambda_2+10\lambda_3)=0$.
The entropy functional proposed for this action is \cite{solo}
\be \label{area}
S_{EE}=\frac{2\pi}{\lp^{3}}\int d^{3}x \sqrt{h}(1+\frac{L^{2}}{{2}}(2\lambda_{3}R+\lambda_{2}( R_{\mu\nu}n^{\nu}_{i}n^{\mu}_{i}-\frac{1}{2} \K^{i}\K_{i})+2\lambda_{1} (R_{\mu\nu\rho\sigma}n^{\mu}_{i}n^{\nu}_{j}n^{\rho}_{i}n^{\sigma}_{j}-\K^{i}_{ab}\K^{ab}_{i}))\,.\\
\ee 
Here $i$ denotes the two transverse directions $\rho=f(z)$ and $ \tau=0$ and $\K_{i}$'s are the two extrinsic curvatures along these two directions pulled back to the surface. The extrinsic curvature for $n_{\tau}$ is zero.
We have to minimize this entropy functional to determine how the entangling surface probes the bulk spacetime. We put $\rho=f(z), \tau=0$ in the metric and minimize  (\ref{area}) on this codimension 2 surface and find the Euler-Lagrange equation for $f(z)$. Using the solution for $f(z)$ we evaluate (\ref{area}) to get the EE.  

\noindent For  the sphere, we get  $f(z)=\sqrt{f_{0}^{2}-z^{2}}$ which gives the universal log  term,
\be
S_{EE}= -4 a \ln(\frac{f_{0}}{\delta})\,. 
\ee
For the cylinder, $f(z)=f_{0}-\frac{z^{2}}{4f_{0}}+...$ which gives,
\be
S_{EE}= - \frac{c H}{2 R} \ln(\frac{f_{0}}{\delta})\,. 
\ee
 \be \label{ac} a= \frac{\pi^{2} L^{3}}{f_{\infty}^{3/2}\lp^{3}}(1-2f_{\infty}(\lambda_{1}+2\lambda_{2}+10\lambda_{3}))\,\, {\rm {and}}\,\,  c= \frac{\pi^{2} L^{3}}{f_{\infty}^{3/2}\lp^{3}}(1+2f_{\infty}(\lambda_{1}-2\lambda_{2}-10\lambda_{3}))\,.\ee
and $\delta$ is the UV cut-off comes from the lower limit of the $z$ integral. $f_{0}$ is the radius of the entangling surface. These are the expected results \cite{higher, gbholo}.

\section{Generalized Gravitational Entropy} 
Following  \cite{maldacena}, the generalized  gravitational entropy is defined as,
\be \label{EE}
S=-n\partial_{n}(\ln[Z(n)]-n\,\ln[Z(1)])_{n=1}\,,
\ee
where $\ln[Z(1)]$ is identified with the Euclidean gravitational action  for which the period of the Euclidean time is $2\pi$ and the boundary condition for other fields collectively denoted as $\phi$ present in the action is $\phi(0)=\phi(2\pi)\,.$ $\ln[Z(n)]$ is identified with the Euclidean gravitational  action $I_{n}$ for which the period of the Euclidean time is $2\pi n$ and the boundary condition for  $\phi$ is still $\phi(0)=\phi(2\pi)\,.$ This is  the usual replica trick. Translating this fact for the holographic case we can define $I_{n}$ for a regularized geometry on a cone whose opening angle is $2\pi/ n$.  We can analytically continue this for non integer $n$ and then can compute the entropy. Also while evaluating $\ln[Z(n)]$ we can perform the time $\tau$ integral from 0 to $2\pi$ and multiply it by $n$ so that $\ln[Z(n)]= n\, \ln[Z]_{2\pi}$\,. The entropy calculated using this method is equal to the  area of  some codimension 2 surface where the time circle shrinks to zero which can be shown to be the minimal surface in Einstein gravity\cite{maldacena}. {
 In this section we will show that this  procedure  also gives the correct entanglement entropy for higher curvature gravity theories.  To compute the EE  we have to start with some specific boundary geometry for the $n^{th}$  solution. Then we can construct our bulk spacetime using the  Fefferman-Graham expansion.
We will consider the following two 4-dimensional metrics following \cite{solo},
\begin{align} \label {metric1}
\begin{split}
ds^{2}_{cylinder}&= f(r,b) dr^{2}+r^{2} d\tau^{2}+({  f_{0}}+r^{n} d^{1-n}\,\cos(\tau))^{2}d\phi^{2}+dz^{2}\\
ds^{2}_{sphere}&=f(r,b) dr^{2}+r^{2} d\tau^{2}+( { f_{0}}+ r^{n}d^{1-n}\,\cos(\tau))^{2}(d\theta^{2} +\sin^{2}\theta d\phi^{2})\\
\end{split}
\end{align}
where, $f(r,b)=\frac{r^{2}+b^{2} n^{2}}{r^{2}+b^{2}}\,.$ For $b\rightarrow 0$ and $ n\rightarrow 1$ limit these two metrics reduce to the  cylinder and  the  sphere. The key point in eq.(\ref{metric1}) as compared to earlier regularizations e.g., \cite{maldacena} is the introduction of a regulator in the extrinsic curvature terms. This is needed since otherwise the Ricci scalar would go like $(n-1)/r$ and would be singular. Another important point is that $b$ is a regulator which at this stage does not have an restriction except that $f(0,b)=n^2$. In AdS/CFT we do not expect an arbitrary parameter to appear in the metric.  $b$ is here a dimensionful quantity having the  dimension of $r\,.$  So $b$ must be  proportional to  $f_{0} (n-1)^{\alpha>0}$ such that it goes to zero as $n\rightarrow 1$. We can take the metrics in eq.(\ref{metric1}) as boundary metrics and construct the bulk spacetime using the Fefferman-Graham expansion.  Notice that our starting point is a smooth metric. At the end of the calculation, when we remove the regulators and compute EE, we will separately check what the contribution from the singularities is going to be. In the best case scenario, although the boundary metric will be singular once the regulator is removed, the bulk metric will at most be mildly singular, namely the on-shell bulk action will not be singular, following the terminology used in \cite{maldacena}. As in \cite{maldacena} we could have done a conformal transformation to pull out a factor of $r^{2}$ such that the $r, \tau$ part of the metric  looks like $ d\tau^{2}+\frac{dr^2}{r^{2}}$ which would make the time-circle non-shrinking.  We can use this form of the metric with  a suitable regularization and do the calculation after verifying that there are no singularities  in the bulk. Since this is a conformal transformation of a smooth metric, the  results for the universal part of the EE will remain unchanged. One can write the bulk metric as,
\be \label{bulk}
ds^{2}= \tilde L^2\frac{d\rho^{2}}{4\rho^2}+\frac{( g^{(0)}_{ij}+\rho g^{(2)}_{ij}+.....)}{\rho} dx^{i}dx^{j}\,.\\
\ee
To evaluate the $\log$ term we will need the $g^{(2)}_{ij}$ coefficient and here we will use eq.(\ref{metric1}) as  $g^{(0)}_{ij}\,.$  We will consider here a 5 dimensional bulk lagrangian.
In this case, $$ g^{(2)}_{ij}= -\frac{\tilde L^{2}}{2}(R^{(0)}_{ij}-\frac{1}{6}g^{(0)}_{ij} R^{(0)})\,,$$ where $R^{(0)}_{ij}$ and $R^{(0)}$ are constructed using $g^{(0)}_{ij}\,.$  Note that in  all  subsequent calculations $g^{(2)}_{ij}$ will play an important role.  The structure of $g^{(2)}_{ij}$ is independent of the form of the higher derivative terms present in the action. Only terms proportional to $n-1$ in the on-shell bulk action contributes to the $S_{EE}.$  The calculation is similar in spirit to the way that Weyl anomaly is extracted in AdS/CFT, e.g., \cite{gbholo} except that the $n-1$ dependence comes from the neighbourhood of $r=0$ in the bulk action. In the next section we proceed to give details of this.} 


\subsection*{Regularization procedure}
To illustrate the regularization procedure in some detail, we start with some simple examples involving  curvature polynomials \footnote{We thank Sasha Patrushev for discussions on this topic.}.  We calculate $g^{(2)}_{ij}$ and evaluate  the following integral ,
\be
I_{1}=\int d^{5} x\,\sqrt{g}\, R_{\mu\nu} R^{\mu\nu} \,.\\
\ee
Following\footnote{Alternatively we could have done the expansion around $x=0$ first, since it was assumed in \cite{solo} that the metric is valid between $0<r<b\ll  f_{0}$. Then we could have integrated $x$ in the neighbourhood of $x=0$. The results are identical.} \cite{solo}, in the integrand, we put $r= b x$  then expand around $b=0$ and pick out the $O(b^0)$ term. The $r$ integral is between $0<r<r_0$. This makes the upper limit of the $x$ integral to be $r_0/b$ which goes to infinity.
We will be interested in the log term so we extract first the coefficient of $\frac{1}{\rho}$ term which has the following form,
\be
I_{1}= b \int \frac{d\rho}{\rho}\, d\tau\,d^{2}y \int_{0}^{\infty} dx\,(n-1)^{2}\zeta(x, n) (bx)^{2n-3}+O((n-1)^3)\,.
\ee
 We have here shown only the leading term. Note that at this stage the integrand is proportional to $(n-1)^2$ whereas we need get something proportional to $(n-1)$. The integral over $x$ will give a factor of $1/(n-1)$. We will now expand $\zeta(x,n)$ around $n=1$ and then carry out the integral over $x$. After expanding around $n=1$ this leads to
\be
I_{1}= (n-1) \zeta_{1}+ \mathcal{O}(n-1)^{2}+\cdots\,.
\ee
Note that the $r^{n}$ factor in the cylindrical and the spherical  parts in (\ref{metric1}) were crucial in reaching this point.
$\zeta_{1}$ is just a quantity independent of the regularization parameters $b$, $d$, $\epsilon$, $\epsilon'$. 
The same procedure is applied for other curvature polynomial integrals. For example, 
\begin{align}
\begin{split}
I_{2}&=\int \sqrt{g}\, d^{5}x R_{\mu\nu\rho\sigma}R^{\mu\nu\rho\sigma}= (n-1)\zeta_{2}+\mathcal{O}(n-1)^{2}+\cdots\,,\\
I_{3}&=\int \sqrt{g}\, d^{5} x R^{2}= \mathcal{O}(n-1)^{2}+\cdots\,.\\
\end{split}
\end{align}

\subsection {Four derivative theory}

Let us now consider the general $R^{2}$ theory lagrangian action given in eq.(\ref{R})\,. Also we will henceforth consider only a 5 dimensional bulk spacetime unless mentioned otherwise. The boundary of this spacetime  is at $\rho=0\,.$  We then evaluate the  total action and extract the $\frac{1}{\rho}$ term and carry out the $\tau$ integral. We put $r= b\, x$  and expand (\ref{R}) around $b=0\,.$ Then we  pick out the $O(b^0)$ term.
\be \label{int}
S=-\frac{1}{2\lp^{3}} \int \,{\frac{d\rho}{\rho}}\, dx\, d^{2}y\, (n-1)^{2} a_{1} \frac{(b x)^{2n}}{x^{3}}+\mathcal{O}((n-1)^{3})\,,\\
\ee
where
\be 
a_{1}= \frac{ A(x)}{18\, b^2 f_{\infty}^{5/2}{f_{0}}\,(1+x^{2})^{4}}\,.
\ee
$A(x)$ is a function of $x\,.$ For the cylinder we get,
\begin{align}
\begin{split}
A(x)&=\pi  L^3   \big(f_{\infty}^2 \left(\lambda_{1} \left(4 x^8+16 x^6+43 x^4+36 x^2+9\right)-2 \left(20 x^8+80 x^6+161 x^4+108 x^2+27\right) (\lambda_{2}+5 \lambda_{3})\right)\\&+6 f_{\infty} \left(5 x^8+20 x^6+38 x^4+24 x^2+6\right)-3 \left(8 x^8+32 x^6+59 x^4+36 x^2+9\right)\big)\,.
\end{split}
\end{align}
 We then carry out the $x$ integral.
\be \label{int1}
S=-\frac{1}{2\lp^{3}}\int \,{\frac{d\rho}{\rho}}\, d^{2}y\,\frac{A_{1}(x,n)}{36\, b^2\, f_{\infty}^{5/2} \left(n^2-1\right) { f_{0}}\, x^2}{\bigg |}_{0}^{\infty}\,,\\
\ee
where
\begin{align}
\begin{split}
A_{1}(x,n)&=\pi  L^3 (n-1)^2  (b x)^{2 n}\bigg{[}(n-1) x^4 \, _2F_1\left(2,n+1;n+2;-x^2\right) \left(f_{\infty}^2 (5 \lambda_{1}-14 (\lambda_{2}+5 \lambda_{3}))+6 f_{\infty}-3\right)\\&+2 (n-1) x^4 \, _2F_1\left(3,n+1;n+2;-x^2\right) \left(f_{\infty}^2 (5 \lambda_{1}-14 (\lambda_{2}+5 \lambda_{3}))+6 f_{\infty}-3\right)\\&+ \,_2F_1\left(4,n+1;n+2;-x^2\right)\big(4 f_{\infty}^2 \lambda_{1} x^4 (1-n ) -40 f_{\infty}^2 \lambda _{2} x^4(1- n ) -200 f_{\infty}^2 \lambda _{3} x^4 (1- n )\\&+30f_{\infty}x^{4}(1-n)-24 x^{4}(1-n)\big )-9 f_{\infty}^2\lambda_{1}(1+n)+54 f_{\infty}^2 \lambda_{2}(1+n)+270 f_{\infty}^2 \lambda_{3}(1+n)\\& -36 f_{\infty}(1+ n)+27(1+ n)\bigg{]}\,.
\end{split}
\end{align}
For the cylinder after doing the expansion around $n=1$ and  the remaining integrals (note that $\rho=z^2$ in the coordinates used in \cite{higher} and so $\ln \delta_{\rho}=2\ln \delta$),
\be
S_{EE}=-\frac{c H}{2 R} \ln(\frac{{ f_{0}}}{\delta})\,.\\
\ee
Here we have used $1=f_\infty-\frac{1}{3}f_\infty^2(\lambda_1+2\lambda_2+10\lambda_3)$ and $c$ is given in eq.(\ref{ac}).
For the sphere we proceed similarly. In this case, expanding (\ref{R})  around $b=0$ we get ,
\be \label{int2}
S=\cdots-\frac{1}{2\lp^{3}} \int \,{\frac{d\rho}{\rho}}\, dx\, d^{2}y\, (n-1)^{2} a_{1} \frac{(b x)^{2n}}{x^{3}}+\mathcal{O}((n-1)^{3})\,,\\
\ee
where
\be 
a_{1}= \frac{ A(x)}{72\, b^2 f_{\infty}^{5/2} { f_{0}^{4}}\,(1+x^{2})^{4}}\,.
\ee
$A(x)$ is a function of $x\,.$ For the sphere we get,
\begin{align}
\begin{split}
A(x)&=-\pi  L^3 \sin (\theta ) \bigg{[}300\, b^4 \lambda _3 x^{10} f_{\infty }^2-45\, b^4 x^{10} f_{\infty }+600 \,b^4 \lambda _3 x^8 f_{\infty }^2-90\, b^4 x^8 f_{\infty }+300\, b^4 \lambda _3 x^6 f_{\infty }^2\\&-45\, b^4 x^6 f_{\infty }+36\, b^4 x^{10}+72\, b^4 x^8+36\, b^4 x^6-680\, b^2 \lambda _3 R^2 x^{10} f_{\infty }^2+84\, b^2 R^2 x^{10} f_{\infty }\\&-1920\, b^2 \lambda _3 R^2 x^8 f_{\infty }^2+216\, b^2 R^2 x^8 f_{\infty }-120\, b^2 \lambda _3 R^2 x^6 f_{\infty }^2+36\, b^2 R^2 x^6 f_{\infty }+1120\, b^2 \lambda _3 R^2 x^4 f_{\infty }^2\\&-96\, b^2 R^2 x^4 f_{\infty }-60\, b^2 R^2 x^{10}-144\, b^2 R^2 x^8-36\, b^2 R^2 x^6+48\, b^2 R^2 x^4+2 \lambda _1 f_{\infty }^2 \big(-3 b^4 \left(x^2+1\right)^2 x^6\\&+2 b^2 R^2 \left(7 x^6+24 x^4-3 x^2-20\right) x^4+4 R^4 \left(x^8-73 x^6+242 x^4+361 x^2+54\right)\big)\\&+4 \lambda _2 f_{\infty }^2 \big(15 b^4 \left(x^2+1\right)^2 x^6-2 b^2 R^2 \left(17 x^6+48 x^4+3 x^2-28\right) x^4+4 R^4 \big(13 x^8-13 x^6\\&+230 x^4+301 x^2+54\big)\big)+4320 \lambda _3 R^4 f_{\infty }^2+1040 \lambda _3 R^4 x^8 f_{\infty }^2-192 R^4 x^8 f_{\infty }-1040 \lambda _3 R^4 x^6 f_{\infty }^2\\&-168 R^4 x^6 f_{\infty }+18400 \lambda _3 R^4 x^4 f_{\infty }^2-2760 R^4 x^4 f_{\infty }+24080 \lambda _3 R^4 x^2 f_{\infty }^2-3144 R^4 x^2 f_{\infty }-576 R^4 f_{\infty }\\&+168 R^4 x^8+264 R^4 x^6+2208 R^4 x^4+2328 R^4 x^2+432 R^4)\bigg{]}\,.
\end{split}
\end{align}
After doing the $x$ integral,
\be 
S=-\frac{1}{2\lp^{3}}\int \,{\frac{d\rho}{\rho}}\, d^{2}y\,\frac{A_{1}(x,n)}{144\, b^2\, f_{\infty}^{5/2}\,n\, \left(n+1\right){ f_{0}^{4}}\, x^2}{\bigg |}_{0}^\infty\,,\\
\ee
where $A_{1}(x,n)$ is a function of $x$ and $n\,.$
\begin{align}
\begin{split}
A_{1}(x,n)&=\pi  L^3 (n-1) \sin (\theta ) (b x)^{2 n} \bigg{[}-8 (n+1) R^4 \big(f_{\infty }^2 (\lambda _1 (145  x^2(n-1)+54 n)+2 (\lambda _2+5 \lambda _3)\\& (85  x^2(n-1)+54 n))-3 f_{\infty } (n (35 x^2+24)-35 x^2)+n \left(75 x^2+54\right)-75 x^2\big)\\&+\, _2F_1\left(4,n+1;n+2;-x^2\right)(-72 (n-1) n R^4 x^4 \left(\left(\lambda _1+2 \left(\lambda _2+5 \lambda _3\right)\right) f_{\infty }^2-3 f_{\infty }+3\right))\\&+\, _2F_1\left(3,n+1;n+2;-x^2\right)(8 (n-1) n R^2 x^4 (f_{\infty }^2 (\lambda _1 (15 b^2+328 R^2)-2 \left(\lambda _2+5 \lambda _3\right) \\&\left(21 b^2-232 R^2\right))+6 \left(3 b^2-46 R^2\right) f_{\infty }-9 b^2+192 R^2))+\, _2F_1\left(1,n+1;n+2;-x^2\right)\\&((n-1) n x^4 (-36 b^4+60 b^2 R^2+f_{\infty } (45 b^4-84 b^2 R^2+2 f_{\infty } (\lambda _1 (3 b^4-14 b^2 R^2+580 R^4)\\&-2 (\lambda _2+5 \lambda _3) (15 b^4-34 b^2 R^2-340 R^4))-840 R^4)+600 R^4))+\, _2F_1\left(2,n+1;n+2;-x^2\right)\\&(-3 (n-1) n x^4 (2 f_{\infty }^2 (\lambda _1 (b^4+2 b^2 R^2-264 R^4)-2 (\lambda _2+5 \lambda _3) (5 b^4-2 b^2 R^2+168 R^4))\\&+3 (5 b^4-4 b^2 R^2+136 R^4) f_{\infty }-12 (b^4-b^2 R^2+24 R^4)))\bigg{]}\,.\\
\end{split}
\end{align}
For the sphere after doing the expansion around $n=1$ and  the remaining integrals ,
\be
S_{EE}=-4\,a\, \ln(\frac{{f_{0}}}{\delta})\,,\\
\ee
where we have used  $1=f_\infty-\frac{1}{3}f_\infty^2(\lambda_1+2\lambda_2+10\lambda_3)$  and $a$ is given in eq.(\ref{ac}). Thus we get the expected universal terms using the regularization proposed in \cite{solo}.
\subsection {New Massive Gravity}
As an example for a calculation of generalized gravitational entropy in other dimensions, we consider the New Massive Gravity  action in three dimensions \cite{townsend} and use the notation in \cite{AS} $$ S =-\frac{1}{2\lp}\int d^{3} x\sqrt{g}\big[R+\frac{2}{L^{2}}+ 4\lambda L^{2}(R_{ab}R^{ab}-\frac{3}{8} R^{2})\big]\,.$$ 
Here $1-f_\infty+f_\infty^2 \lambda=0$. 
The entropy functional for this is not intrinsic as compared to the three dimensional Einstein gravity and is given by  \be S_{EE}=\frac{2\pi}{\lp}\int dx\,\sqrt{g_{xx}} \,\big[1+4 \lambda L^{2}([ R_{\mu\nu}n^{\mu}_{i}n^{\nu}_{i}-\frac{1}{2}K^{i}K_{i}]-\frac{3}{4} R)\big]\,.\ee
The integral is over the one dimensional entangling region. 
We calculate the generalized gravitational entropy following the same procedure as used above. The two dimensional squashed cone metric is given by 
$$
ds^{2}= f(r,b) dr^{2}+r^{2} d\tau^{2}\,.
$$
$f_{0}$ in this case also corresponds to the radius of the entangling surface.\par  In 3 dimensions \cite{skenderis,skenderis1}
\be
g^{(2)}_{ij}=-\frac{\tilde L^{2}}{2} R^{(0)}g^{(0)}_{ij}+t_{ij}
\ee
Only divergence and trace of $t_{ij}$ are known.
$$ g^{(0)}_{ij}t^{ij}= R^{(0)}\,,\quad \nabla^{i}t_{ij}=0\,.$$
\be
R^{(0)}=-\frac{2 b^2 \left(n^2-1\right)}{\left(b^2 n^2+r^2\right)^2}\,.\\
\ee
 Using $1-f_{\infty}+f_{\infty}^{2}\lambda=0$ and we get,
\be
S=\cdots+\frac{1}{2\lp}\int {\frac{d\rho}{\rho}}\,\int _{0}^{2\pi}d\tau\,\int_{r=0}^{r={ f_{0}}}dr\, \frac{L (r b^2 (n^2-1) (1+2f_{\infty}\lambda)}{ f_{\infty}^{1/2} \sqrt{b^2+r^2} (b^2 n^2+r^2)^{3/2}}+\cdots\,.
\ee
Note that $t_{ij}$ does not enter in the calculation of the universal term. After doing the integrals we get
\be
S=\cdots+\int \frac{d\rho}{\rho}\big[\frac{ \pi  L \left(1+2  f_{\infty }\lambda \right)}{\lp\sqrt{f_{\infty }}}\left(\frac{1}{n}-\sqrt{\frac{b^2+{ f_{0}}^2}{b^2 n^2+{f_{0}}^2}}\right)\big]+\cdots\,.
\ee
Then expanding around $b=0$ and $n=1$  we get the correct universal term
\be
S_{EE}= \frac{c}{3}\ln(\frac{{f_{0}}}{\delta})\,,\\
\ee
where, $\frac{c}{3}= \frac{2\pi L (1+2f_{\infty}\lambda)}{f_{\infty}^{1/2}\lp}\,.$\subsection {Quasi-Topological Gravity}
The six-derivative action for quasi-topological gravity is given below \cite{quasi},
\begin{align}
\begin{split} \label{qtop}
S =-\frac{1}{2\lp^{3}}\int d^{5} x\sqrt{g}\big[R+\frac{12}{L^{2}}+\frac{L^{2}\lambda}{2}GB+\frac{L^{4} 7\mu}{4}Z_{5}\big]
\end{split}
\end{align}
where, \begin{align}
\begin{split} GB&= R_{\mu\nu\rho\sigma}R^{\mu\nu\rho\sigma}-4 R_{\mu\nu}R^{\mu\nu}+R^{2}  \,\,\rm{and}\\ Z_{5}&=R_{\mu}{}^{\nu}{}_{\rho}{}^{\sigma}R_{\nu}{}^{\alpha}{}_{\sigma}{}^{\beta}R_{\alpha}{}^{\mu}{}_{\beta}{}^{\rho}+\frac{3}{8} R_{\mu\nu\rho\sigma}R^{\mu\nu\rho\sigma}R-\frac{9}{7} R_{\mu\nu\rho\sigma}R^{\mu\nu\rho}{}_{\alpha}R^{\sigma\alpha} +\frac{15}{7}R_{\mu\nu\rho\sigma}R^{\mu\rho}R^{\nu\sigma}\\&+\frac{18}{7}R_{\mu\sigma}R^{\sigma\alpha}R^{\mu}{}_{\alpha}-\frac{33}{14} R_{\alpha\beta}R^{\alpha\beta}R+\frac{15}{56} R^{3}\,.\\
\end{split}
\end{align}
Following exactly the same procedure  we can derive the holographic entanglement entropy for this six derivative gravity theory.\\
For the sphere we get,
\be
S_{EE}=-\frac{4\pi^{2}  L^{3}}{f_{\infty}^{3/2}\lp^{3}}(1- 6 f_{\infty}\lambda +9 f_{\infty}^{2}\mu)\ln(\frac{{ f_{0}}}{\delta})\,.\\
\ee
For the cylinder
\be
S_{EE}=-\frac{\pi^{2}  L^{3}H}{2 f_{\infty}^{3/2} \lp^{3}R}(1- 2 f_{\infty}\lambda -3 f_{\infty}^{2}\mu)\ln(\frac{{ f_{0}}}{\delta})\,.\\
\ee
These are the  correct  universal terms.
\subsection{$\alpha'^3$  IIB supergravity}
The action for this follows from \cite{r4}
\begin{align} \label{Weyl4}
\begin{split}
S =-\frac{1}{2\lp^{3}}\int d^{5} x\sqrt{g}\big[R+\frac{12}{L^{2}}+L^{6} \gamma \kappa_{5}\big]
\end{split}
\end{align}
where, $$\kappa_{5}=C_{\alpha\beta \mu\nu}C^{\rho\beta \mu\sigma}C^{\alpha\delta\gamma}{}_{\rho}C^{\nu}{}_{ \delta\gamma\sigma}-\frac{1}{4}C_{\alpha\beta \mu\nu}C^{\alpha\beta}{}_{ \rho\sigma}C^{\mu\rho}{}_{\delta\gamma}C^{\nu\sigma\delta\gamma}\,.$$
$C_{\alpha\beta\mu\nu}$ is the Weyl tensor in 5 dimensions. In the context of IIB string theory, $\gamma=\frac{1}{8}\zeta(3)\alpha'^3/L^6$.
For this theory we find that the universal parts of EE do not get corrected compared to the Einstein case. This is expected since from the perspective of the AdS/CFT correspondence, the $C^4$ correction correspond to $1/\lambda$ corrections and the anomalies are not expected to receive such corrections. Recently the effect of the $C^4$ correction on Renyi entropy was analysed in \cite{galante}.

\subsection{Comment about singularities in the metric}
 There are singularities in the five dimensional metric coming entirely from $g^{(2)}_{ij}\,.$ We expand the metric around $r=0\,.$ Upto the  leading order the metric is shown below. 

\noindent For the sphere (diagonal components are $g_{\rho\rho},g_{rr},g_{\tau\tau}, g_{\theta\theta},g_{\phi\phi}$ ), 
\be
\left(
\begin{array}{ccccc}
 \frac{L^2}{4 f_{\infty} \rho ^2} & 0 & 0 & 0 & 0 \\
 0 & \frac{(n-1) \cos (\tau ) L^2}{ { f_{0}}\, r f_{\infty }}+\frac{1}{\rho } & 0 & 0 & 0 \\
 0 & 0 & \frac{r^{2}}{\rho}-\frac{ L^2 (n-1)\, r \cos (\tau )}{ {f_{0}} f_{\infty }} & 0 & 0 \\
 0 & 0 & 0 & \frac{{ f_{0}}^2}{\rho } & 0 \\
 0 & 0 & 0 & 0 & \frac{{ f_{0}}^2 \sin ^2(\theta )}{\rho } \\
\end{array}
\right)\,.
\ee
For the cylinder,
\be
\left(
\begin{array}{ccccc}
 \frac{L^2}{4 f_{\infty} \rho ^2} & 0 & 0 & 0 & 0 \\
 0 & \frac{(n-1) \cos (\tau ) L^2}{2 { f_{0}}\, r f_{\infty }}+\frac{1}{\rho } & 0 & 0 & 0 \\
 0 & 0 & \frac{r^{2}}{\rho}-\frac{ L^2 (n-1)\, r \cos (\tau )}{2 { f_{0}} f_{\infty }} & 0 & 0 \\
 0 & 0 & 0 & \frac{{ f_{0}}^2}{\rho } & 0 \\
 0 & 0 & 0 & 0 & \frac{1}{\rho } \\
\end{array}
\right)\,.
\ee
The $g_{rr}$ component is singular in $r$.
The other components are non  singular.  However it is easy to see that the determinant does not have a singularity at $r=0$.  The singularity in the metric  gives rise to singularities in the components of the Riemann tensor. We have explicitly checked that  these singularities do not enter in the higher derivative actions considered in this paper. Hence these are mild singularities in the sense used in \cite{maldacena}. Note that in order to calculate  the universal part of EE  in four dimensions  only $g^{(2)}_{ij}$ is important. 
\section{ Wald Entropy  }
In this section we turn to the computation of Wald entropy for the higher derivative theories considered above. We will compute the Wald entropy on the surface $r=0=\tau$. The reason for this will become clear shortly.
\subsection{Four derivative theory}
The Wald entropy calculated from eq(\ref{R}) is given by
\be \label{wald} S_{wald}=\int d^{d-2}x \sqrt{h}\frac{\partial L}{\partial R_{\alpha\beta\gamma\delta}} \hat \epsilon_{\alpha\beta}\hat \epsilon_{\gamma\delta}\,.\\
\ee
This expression is evaluated on a codimension-2 surface. Here $\hat \epsilon_{\alpha \beta}=n_{\alpha}^{1}n_{\beta}^{2}-n_{\alpha}^{2}n_{\beta}^{1}$ is the binormal corresponding to the two transverse directions $1,2\,.$  For the four derivative theory,
\begin{align}
\begin{split}
\frac{\partial L}{\partial R_{\alpha\beta\gamma\delta}}&=\frac{1}{2}(g^{\alpha\gamma}g^{\beta\delta}-g^{\alpha\delta}g^{\beta\gamma})+L^{2} \big[\lambda_{1}R^{\alpha\beta\gamma\delta}+\frac{1}{4} \lambda _{2} \left(g^{\beta \delta } R^{\alpha \gamma }-g^{\beta \gamma } R^{\alpha \delta }-g^{\alpha \delta } R^{\beta \gamma }+g^{\alpha \gamma } R^{\beta \delta }\right)\\& +\frac{1}{2}\lambda _3 R \left(g^{\alpha \gamma } g^{\beta \delta }-g^{\alpha \delta } g^{\beta \gamma }\right)\big]\,.\\
\end{split}
\end{align}
Then after some simplifications we get,
\be \label{waldf}
S_{wald}= \frac{2\pi}{\lp^{3}}\int d^{3}x \sqrt{h}\big(1+\frac{L^{2}}{{2}}(2\lambda_{3}R+\lambda_{2} R_{\mu\nu}n^{\nu}_{i}n^{\mu}_{i}+2\lambda_{1} R_{\mu\nu\rho\sigma}n^{\mu}_{i}n^{\nu}_{j}n^{\rho}_{i}n^{\sigma}_{j})\big)\,.\\
\ee
  In this section we will show that starting with the boundary metrics  in  eq.(\ref{metric1}) we  can construct a bulk spacetime on which $S_{wald}$ will produce the expected universal parts for the entanglement entropy for both cylinder and sphere. Note that (\ref{waldf}) differs from (\ref{area}) by the $O(\K^{2})$ terms.

\subsection*{Cylinder}
As we will show,  a particular form of the regularization $b=\alpha (n-1)^{1/2}$, where $\alpha$ is some number which we will determine later (it will turn out to be surface dependent but theory independent), is needed to get the correct universal term. Recall that the only restriction on $b$ was that $f(r,b)$ has to be $ n^2$ in the $r=0$ limit. However, in holographic calculations we expect that the bulk metrics will only depend on the AdS radius, the radius of the entangling region and $n$. As such we can expect that the only way that $b\rightarrow 0$ would arise in holographic calculations is such that $b$ is some positive power of $(n-1)$.  Now we will evaluate eq.(\ref{waldf}) using eq.(\ref{bulk}) using the cylinder metric to be its boundary. Then we extract  the coefficient of the $\frac{1}{\rho}$ term. We set $\tau =0\,.$ There is no integral over $r$ in the Wald entropy as the entangling surface is located at $r=0, \tau=0\,.$ We put $r= b\, x\,.$  After that we expand around $x=0$ and then expand around $n=1\,.$ We retain only the $n$ independent part as other terms vanish in $n\rightarrow 1$ limit.   Below we quote some intermediate steps after expanding in $\rho$, $r$ and $n$ respectively. It is important to take the limits in $r$, $n$ in that particular in order to get the correct result \cite{solo}.  After doing the $\rho$ expansion we pick out the $\frac{1}{\rho}$ term of (\ref{waldf}) which is shown below.
\be \label{waldeq}
S_{wald}=\cdots+ \frac{2\pi}{\lp^{3}}\int d\rho d \phi dz \frac{ A(x,n)}{\rho}+\mathcal{O}(\rho)+\cdots\,,
\ee
where $$A(x,n)=\frac{L^{3} \left(n^2-1\right) d^{-n} \left(  \left(4 \lambda _2+20 \lambda _3-2\lambda_{1}\right) f_{\infty }-1\right) \left(2 { f_{0}} d^n-d \left(n^2+n+x^2-2\right) (b x)^n\right)}{24 b^2 f_{\infty }^{3/2} \left(n^2+x^2\right)^2}\,.$$
Then expanding $A(x,n)$ around $x=0$  we get,
\be \label{ren2}
A(x,n)=\frac{L^3 \left(n^2-1\right) { f_{0}} \left(\left(4\lambda _2+20 \lambda _3-2\lambda _1\right) f_{\infty }-1\right)}{12\, b^2 n^4 f_{\infty }^{3/2}}+\cdots\,.
\ee
If $$b=\frac{2 { f_{0}}}{\sqrt{3}}\sqrt{n^2-1}\beta(n)\,,$$ 
where $\beta(1)=1$ we get upon further expanding $A(x,n)$ around $n=1$
\be
A(x,n)=-\frac{L^3 \left(1+2 \left(\lambda _1-2 \left(\lambda _2+5 \lambda _3\right)\right) f_{\infty }\right)}{16 {f_{0}} f_{\infty }^{3/2}}+\mathcal{O}(n-1)+\cdots\,.
\ee
 Notice that the choice for $b$ was independent of the theory, i.e., in this case of $\lambda_i$'s. Finally we get,
\be
S_{wald}= -\frac{ \pi ^2  L^{3} H (1+2 f_{\infty} (\lambda_{1}-2 \lambda_{2}-10 \lambda_{3}))}{2 f_{\infty}^{3/2}\lp^{3} { f_{0}} }\ln(\frac{{ f_{0}}}{\delta})\,.
\ee
This is precisely what is expected.
\subsection*{Sphere}
We proceed similarly for the sphere case. First we expand in $\rho$ and pick out the $\frac{1}{\rho}$ term.
\be \label{waldeq}
S_{wald}=\cdots+ \frac{2\pi}{\lp^{3}}\int d\rho d \theta d\phi \frac{ A(x,n)}{\rho}+\mathcal{O}(\rho)+\cdots\,.
\ee
Here
\begin{align}
\begin{split}
A(x,n)&=\frac{L^3 d^{-2 n} \sin (\theta )}{12 b^2 f^{3/2} x^2 (n^2+x^2)^2} \big[4 \lambda _1 f_{\infty } (b^2 x^2 d^{2 n} (n^2+x^2)^2-d^2 \, (n^4-n^3 x^2+3 n^2 x^2+n x^2+x^4-x^2) (b x)^{2 n}\\&+d\, (n^2-1) R\, x^2 (n^2+n+x^2-2) (b\, d x)^n-(n^2-1) R^2 x^2 d^{2 n})-(2 (\lambda _1+2 (\lambda _2+5 \lambda _3)) f_{\infty }-1)\\& (-2 b^2 x^2 d^{2 n}\, (n^2+x^2)^2+d^2 (n^4 (3 x^2+2)+n^3 x^2+3 n^2 x^4-n x^2-x^4+x^2) (b x)^{2 n}\\&+d\, (n^2-1) R x^2 (n^2+n+x^2-2) (b\, d\, x)^n-(n^2-1) R^2 x^2 \,d^{2 n})\big]\\
\end{split}
\end{align}
Then expanding $A(x,n)$ around $x=0$ we get \footnote{Remember that at this stage $n=1+\epsilon$. Thus we will drop $x^{2n}$ compared to $x^2$.},
\be \label{ren}
A(x,n)=\frac{L^3 \sin (\theta )\big(2 b^2 n^4 \left(4 \left(\lambda _1+\lambda _2+5 \lambda _3\right) f_{\infty }-1\right)+\left(n^2-1\right) {f_{0}}^2 \left(\left(-2 \lambda _1+4 \lambda _2+20 \lambda _3\right) f_{\infty }-1\right)\big)}{12 b^2 f_\infty^{3/2} n^4}\,.\\
\ee
Only the $x$ independent term is shown. If (for consistency checks see below) \be \label{bchoice} b= { f_{0}}\sqrt{n^2-1} \beta(n)\ee where $\beta(1)=1$, expanding around $n=1$ we get,
\be 
A(x,n)=- \frac{L^3 \sin (\theta ) \left(1-2 \left(\lambda _1+2 \left(\lambda _2+5 \lambda _3\right)\right) f_{\infty }\right)}{4 f_\infty^{3/2}}+\mathcal{O}(n-1)+\cdots\,.
\ee
As in the cylinder case, notice that the choice for $b$ is theory independent. Finally we get,
\be
S_{wald}= -\frac{ 4 \pi ^2  L^{3}  (1-2 f_{\infty} (\lambda_{1}+2 \lambda_{2}+10 \lambda_{3}))}{f_{\infty}^{3/2} \lp^{3} }\ln(\frac{{ f_{0}}}{\delta})
\ee 
We have  fixed $b$ for both the cylinder and the sphere case. In all the subsequent calculations of Wald entropy we will use these same values for $b$. 
 \subsection{Quasi-Topological gravity}
The Wald entropy is calculated for (\ref{qtop}) using (\ref{wald})\,. For this case,
\begin{align}
\begin{split}
&\frac{\partial L}{\partial R_{\alpha\beta\gamma\delta}}=\frac{1}{2}(g^{\alpha\gamma}g^{\beta\delta}-g^{\alpha\delta}g^{\beta\gamma})+L^{2} \big[\lambda_{1}R^{\alpha\beta\gamma\delta}+\frac{1}{4} \lambda _{2} \left(g^{\beta \delta } R^{\alpha \gamma }-g^{\beta \gamma } R^{\alpha \delta }-g^{\alpha \delta } R^{\beta \gamma }+g^{\alpha \gamma } R^{\beta \delta }\right)\\& +\frac{1}{2}\lambda _3 R \left(g^{\alpha \gamma } g^{\beta \delta }-g^{\alpha \delta } g^{\beta \gamma }\right)\big]+\frac{7\mu L^{4}}{4}\big[( 3\mu_{1} (R^{\alpha \rho \gamma \sigma}R^{\beta \ \delta}_{\ \rho \ \sigma}-
R^{\alpha \rho \delta \sigma}R^{\beta \ \gamma}_{\ \rho \ \sigma}))+\frac{\mu_{2}}{2}[(g^{\alpha\gamma}g^{\beta\delta}-g^{\alpha\delta}g^{\beta\gamma})R_{\mu\nu\rho\sigma}R^{\mu\nu\rho\sigma}\\&+4\,R\,R^{\alpha\beta\gamma\delta}]+
\frac{\mu_{3}}{4}[g^{\beta \delta } R^{\alpha \rho \sigma \mu } R^{\gamma }{}_{\rho \sigma \mu }-g^{\beta \gamma } R^{\alpha \rho \sigma \mu } R^{\delta }{}_{\rho \sigma \mu }-g^{\alpha \delta } R^{\beta \rho \sigma \mu } R^{\gamma }{}_{\rho \sigma \mu }+g^{\alpha \gamma } R^{\beta \rho \sigma \mu } R^{\delta }{}_{\rho \sigma \mu }\\&-2 R^{\gamma \rho } R^{\alpha \beta \delta }{}_{\rho }+2 R^{\delta \rho } R^{\alpha \beta \gamma }{}_{\rho }+2 R^{\beta \rho } R^{\alpha }{}_{\rho }{}^{\gamma \delta }-2 R^{\alpha \rho } R^{\beta }{}_{\rho }{}^{\gamma \delta }]+\frac{\mu_{4}}{2}(R^{\rho \sigma } [g^{\beta \delta } R^{\alpha }{}_{\rho }{}^{\gamma }{}_{\sigma }-g^{\beta \gamma } R^{\alpha }{}_{\rho }{}^{\delta }{}_{\sigma }\\&-g^{\alpha \delta } R^{\beta }{}_{\rho }{}^{\gamma }{}_{\sigma }+g^{\alpha \gamma } R^{\beta }{}_{\rho }{}^{\delta }{}_{\sigma }]+[R^{\alpha \gamma } R^{\beta \delta }-R^{\alpha \delta } R^{\beta \gamma }])+\frac{3 \mu_{5}}{4}[g^{\beta \delta } R^{\alpha \sigma } R^{\gamma }{}_{\sigma }-g^{\beta \gamma } R^{\alpha \sigma } R^{\delta }{}_{\sigma }\\&-g^{\alpha \delta } R^{\beta \sigma } R^{\gamma }{}_{\sigma }+g^{\alpha \gamma } R^{\beta \sigma } R^{\delta }{}_{\sigma }]+ \frac{\mu_{6}}{2}\big[R \left(g^{\beta \delta } R^{\alpha \gamma }-g^{\beta \gamma } R^{\alpha \delta }+g^{\alpha \gamma } R^{\beta \delta }-g^{\alpha \delta } R^{\beta \gamma }\right)\\&+(g^{\alpha \gamma } g^{\beta \delta }-g^{\alpha \delta } g^{\beta \gamma })R_{\mu \nu }R^{\mu \nu }\big]+\frac{3}{2}\mu_{7}(R^2 [g^{\alpha \gamma } g^{\beta \delta }-g^{\alpha\delta } g^{\beta \gamma }])\big]\,.\\
\end{split}
\end{align}
Now the coefficients are, $$ \mu_{1}=1\,,\mu_{2}=\frac{3}{8}\,,\mu_{3}=-\frac{9}{7}\,,\mu_{4}=\frac{15}{7}\,,\mu_{5}=\frac{18}{7}\,,\mu_{6}=-\frac{33}{14}\,,\mu_{7}=\frac{15}{56}\,,$$
and   $\lambda_2=-4 \lambda_1$, $\lambda_3=\lambda_1=\lambda$.
Proceeding similarly as mentioned for the $R^{2}$ theory we get the expected universal terms.

\noindent For the cylinder \footnote{The $c$ and $a$ coefficients for an arbitrary higher derivative theory can be easily calculated using the short-cut mentioned in the appendix of \cite{sen}.},
\be
S_{wald}=-\frac{\pi^{2}  L^{3}H}{2 f_{\infty}^{3/2} \lp^{3}R}(1- 2 f_{\infty}\lambda -3 f_{\infty}^{2}\mu)\ln(\frac{{ f_{0}}}{\delta})\,.\\
\ee
For the sphere,
\be
S_{wald}=-\frac{4\pi^{2}  L^{3}}{f_{\infty}^{3/2}\lp^{3}}(1- 6 f_{\infty}\lambda +9 f_{\infty}^{2}\mu)\ln(\frac{{ f_{0}}}{\delta})\,.\\
\ee
Again note that the choice for $\alpha$ did not depend on the theory.
\subsection{$\alpha'^3$ IIB supergravity}
The Wald entropy is calculated for (\ref{Weyl4}) using (\ref{wald})\,. For this case,
\begin{align}
\begin{split}
\frac{\partial L}{\partial R_{\alpha\beta\gamma\delta}}&=\frac{1}{2}(g^{\alpha\gamma}g^{\beta\delta}-g^{\alpha\delta}g^{\beta\gamma})+L^{6}\gamma \big[\frac{1}{3}(g^{\beta\gamma}C^{\alpha \mu\delta\nu}C_{\nu\rho\sigma\eta}C_{\mu}{}^{\rho\sigma\eta}-g^{\beta\delta}C^{\alpha \mu\gamma\nu}C_{\nu\rho\sigma\eta}C_{\mu}{}^{\rho\sigma\eta}\\&+g^{\alpha\delta}C^{\beta \mu\gamma\nu}C_{\nu\rho\sigma\eta}C_{\mu}{}^{\rho\sigma\eta}-g^{\alpha\gamma}C^{\beta \mu\delta\nu}C_{\nu\rho\sigma\eta}C_{\mu}{}^{\rho\sigma\eta})+\frac{1}{6}(g^{\alpha\gamma}g^{\beta\delta}-g^{\alpha\delta}g^{\beta\gamma})(C_{\sigma}{}^{\mu}{}_{\nu}{}^{\rho}C^{\sigma \eta\nu \zeta}C_{\eta\rho\zeta\mu}\\&-\frac{1}{2}C_{\mu\nu}{}^{\rho\sigma}C^{\mu\nu\eta\zeta}C_{\eta\rho\zeta\sigma})+\frac{1}{6}( g^{\beta \delta }C^{\alpha \rho \zeta \sigma } C_{\rho \sigma \mu \nu } C^{\gamma }{}_{\zeta }{}^{\mu \nu }- g^{\alpha \delta }C^{\beta \rho \zeta \sigma } C_{\rho \sigma \mu \nu } C^{\gamma }{}_{\zeta }{}^{\mu \nu } -g^{\beta \gamma }C^{\alpha \rho \zeta \sigma } C_{\rho \sigma \mu \nu } C^{\delta}{}_{\zeta }{}^{\mu \nu }\\&+g^{\alpha \gamma }C^{\beta \rho \zeta \sigma } C_{\rho \sigma \mu \nu } C^{\delta }{}_{\zeta }{}^{\mu \nu })+\frac{1}{6}( g^{\beta \delta }C^{\alpha \rho \zeta \sigma } C^{\gamma \mu }{}_{\rho }{}^{\nu } C_{\zeta \sigma \mu \nu }-g^{\alpha \delta }C^{\beta \rho \zeta \sigma } C^{\gamma \mu }{}_{\rho }{}^{\nu } C_{\zeta \sigma \mu \nu }-g^{\beta \gamma }C^{\alpha \rho \zeta \sigma } C^{\delta\mu }{}_{\rho }{}^{\nu } C_{\zeta \sigma \mu \nu }\\&+g^{\alpha\gamma }C^{\beta \rho \zeta \sigma } C^{\delta \mu }{}_{\rho }{}^{\nu } C_{\zeta \sigma \mu \nu })+( C^{\alpha \rho }{}_{\mu }{}^{\sigma }C^{\beta \mu \delta \eta } C^{\gamma }{}_{\rho \eta \sigma }- C^{\beta \rho }{}_{\mu }{}^{\sigma }C^{\alpha \mu \delta \eta } C^{\gamma }{}_{\rho \eta \sigma }- C^{\alpha \rho }{}_{\mu }{}^{\sigma }C^{\beta \mu \gamma \eta } C^{\delta}{}_{\rho \eta \sigma }\\&+ C^{\beta \rho }{}_{\mu }{}^{\sigma }C^{\alpha \mu \gamma \eta } C^{\delta }{}_{\rho \eta \sigma })-\frac{1}{2}(C^{\gamma \delta \sigma \zeta } C^{\beta }{}_{\zeta \mu \rho } C^{\alpha }{}_{\sigma }{}^{\mu \rho }+
C^{\alpha \beta\sigma \zeta } C^{\delta }{}_{\zeta \mu \rho } C^{\gamma }{}_{\sigma }{}^{\mu \rho })+\frac{2}{3}(g^{\alpha \delta }C^{\beta \rho \zeta \nu } C_{\rho \sigma \nu \mu }  C^{\gamma \mu }{}_{\zeta }{}^{\sigma }\\&-g^{\beta \delta }C^{\alpha \rho \zeta \nu } C_{\rho \sigma \nu \mu }  C^{\gamma \mu }{}_{\zeta }{}^{\sigma }+g^{\beta \gamma }C^{\alpha \rho \zeta \nu } C_{\rho \sigma \nu \mu }  C^{\delta \mu }{}_{\zeta }{}^{\sigma }-g^{\alpha \gamma }C^{\beta \rho \zeta \nu } C_{\rho \sigma \nu \mu }  C^{\delta \mu }{}_{\zeta }{}^{\sigma })\big]\,.
\end{split}
\end{align}
Proceeding similarly as mentioned for the $R^{2}$ theory we get the expected universal terms.\par
\noindent For the cylinder,
\be
S_{wald}=-\frac{\pi^{2}  L^{3}H}{2 \lp^{3}R}\ln(\frac{{ f_{0}}}{\delta})\,.\\
\ee
For the sphere,
\be
S_{wald}=-\frac{4\pi^{2}  L^{3}}{\lp^{3}}\ln(\frac{{ f_{0}}}{\delta})\,.\\
\ee
As expected, for this case the universal terms are independent of the higher derivative correction.

\subsection{Connection with Ryu-Takayanagi}
The Ryu-Takayanagi calculation involves the minimization of an entropy functional{\footnote{We thank Rob Myers for discussions on this section.}}.
For both the sphere and the cylinder, one can check that minimizing the Wald area functional in the Fefferman-Graham background for squashed cones leads to the correct universal terms provided we choose $b$ as mentioned above. Recall that 
the Wald entropy functional in AdS spacetime was not the correct one \cite{higher, higher2}. However, our background is not AdS and it turns out that the Wald entropy functional leads to the correct universal terms. We show this for the cylinder, the sphere case working similarly. Putting $r=\mathcal{R}(\rho)=r_0+r_1 \rho^\alpha$ around $\rho=0$ leads to $r_0=0$ and the equation 
$$c n r_1^n \rho^{\alpha n+1}-4 r_1^2 R c^n \alpha(\alpha-2)\rho^{2\alpha}=0\,,$$
where we have shown the leading terms which would contribute around $n=1$. If we set $n=1$  we recover the result $\alpha=1, r_1=-1/(4 \bf f_{0})$ for a cylinder--this is expected. The $n=1$ boundary geometry is just flat space with the dual bulk being AdS. Hence we expect to recover the RT result. However if $n=1+\epsilon$, then it is easy to see that either $r_1=0$ or $\alpha=2$ or $r_1=-1/(4 { f_{0}})$ and $\alpha=1+\epsilon$. As in the RT case, only the linear term in $\mathcal{R}(\rho)$ would have affected the universal term--since $\alpha\neq 1$ if $n=1+\epsilon$ we find that there is no linear term.  For $ n \neq 1$ the minimal surface  is at  $r=0=\tau.$ This is the reason why the Wald entropy on the $r=0=\tau$ surface and the RT entropy functional approach give the same result for the universal terms in the squashed cone background.  We now point out a direct comparison between the calculation done in AdS spacetime and that in the squashed cone background for the sphere in what follows.

The Ryu-Takayanagi prescription was implemented in the following way for a spherical entangling surface. Consider the AdS$_5$ metric with the boundary written in spherical coordinates
\be 
ds^2=\frac{\tilde L^2}{z^2}(dz^2+dt^2+d\hat r^2+\hat r^2 d\theta^2+\hat r^2\sin^2\theta d\phi^2)\,.
\ee
Now put $\hat r=f(z)={f_{0}}+f_2 z^2+\cdots$ and $t=0$ and minimize the relevant entropy functional. Implicitly our analysis says that this surface and the  $r=0=\tau$  surface in the coordinate system we have been using are related. Since in both cases the extrinsic curvatures vanish we can attempt to make a direct comparison. In order to do this we make a coordinate transformation:
\be
\frac{dz}{z}\sqrt{1+f'(z)^2}=\frac{d\rho}{2\rho}\,.
\ee
Around $\rho=0$ we will find $z^2=\rho-2 f_2^2 \rho^2+\cdots$ and $f(z)^2/z^2={f_{0}}^2/\rho+2 { f_{0}} f_2 (1+ {f_{0}} f_2)+\cdots$. Now around $\rho=0$, the metric on the $r=0=\tau$ surface takes the form
\be
ds^2=\tilde L^2 [\frac{d\rho^2}{4\rho^2}+{\mathcal K}(\rho) (d\theta^2+\sin^2\theta d\phi^2)]\,,
\ee
where $${\mathcal K}(\rho)=\frac{{f_{0}}^2}{\rho}-\frac{\tilde L^2}{6 b^2 n^4}(2 b^2 n^4+(n^2-1){ f_{0}}^2)\,.$$  This also shows that  for $n\neq 1$ minimal surface is at $r=0=\tau\,.$
Now choosing $b$ as in eq.(\ref{bchoice}), expanding upto $O((n-1)^0)$ and comparing with the RT calculation we find
$f_2=-1/(2 { f_{0}})$. This is exactly what we would have got if we minimized the RT area functional (or the relevant higher derivative entropy functional) in AdS space. This also serves as a consistency check for the choice of $b$. 

\subsection{Comments on the connection with the Iyer-Wald prescription}
Why does the Wald entropy functional lead to the correct result in our case? Wald's formula in eq.(\ref{wald}) is valid for a surface which is a  local bifurcation surface  on which the Killing field vanishes. For a bifurcation surface, the extrinsic curvatures  vanish.  $S_{EE}$ mentioned in (\ref{area}) differs from $S_{wald}$ only by the extrinsic curvature terms. The Noether charge method of \cite{wald} needs a bifurcation surface to remove various ambiguities \cite{Iyer, Kang}. According to the prescription of Iyer and Wald \cite{Iyer}, in order to compute the entropy for horizons which are not bifurcate, e.g., dynamical horizons, the curvature terms in $\frac{\partial L}{\partial R_{abcd}}$  are replaced by their boost invariant counterparts \cite{Iyer}. To do this we have to construct a boost invariant metric from our original metric. Let $g_{ab}$ be our starting $d$ dimensional metric with the two normals $n_{a}^{1}, n_{b}^{2}\,.$ The boost invariant part of  $g_{ab}$ will only have  terms with the same number of  $n^{1}, n^{2}.$ We then consider a $d-2$ dimensional surface and find a neighbourhood of  it $\mathcal{O}$ such that for  any points $x$ belonging to this neighbourhood, we can  find a point $P$ which lies on a unit affine distance  on a geodesic with a tangent vector $v^{a}$ on the $d-2$ dimensional plane perpendicular to this surface under consideration. Now we assign a coordinate system ${U,V,x_{1},...x_{d-2}}$ for the point $x$ where $U,V$ are the components of $v^{a}$ along $n^{1}_{a}$ and $n^{2}_{a}.$ A change  of normals under the boosts $n^{a}_{1}\rightarrow \alpha n^{a}_{1}, n^{b}_{2}\rightarrow \alpha^{-1} n^{a}_{2}$ will change the coordinates as follows $U \rightarrow \alpha U,V\rightarrow \alpha^{-1} V\,.$
 Now we Taylor expand  $g_{ab}$ around $U$and $V$,
\be
g_{ab}= g^{(0)}_{ab}+ U \partial g +V \partial g + U V \partial \partial g+........... \,.\\
\ee
We have shown  the expansion schematically. Under  boosts,   the terms linear in $U,V$ do not remain invariant. The prescription in \cite{Iyer} is to drop these terms. The $UV$ term is invariant under the  boost. One important point to note is that , $\psi^{a} = U(\frac{\partial }{\partial U})^{a}-V(\frac{\partial }{\partial V})^{a}$ is a Killing field of the metric. This means that  Lie derivative of $g_{ab}$ with respect to $\psi$ is zero. Effectively, we have constructed a new spacetime in which the original dynamical horizon becomes a bifurcate Killing horizon. 


The evidence for the existence of this bifurcation surface would be that extrinsic curvatures for this surface in the bulk background vanishes. Our entangling surface is a codimension-2 surface. Now we calculate the  extrinsic curvatures for this surface in the bulk Fefferman-Graham metric. There will be two of them---one along the direction of the normal $^{(\tau)}n$ for $\tau=0$ and the other one along the normal $^{(r)}n$ for $r=0.$ We start  with the 5 dimensional metrics given in eq.(\ref{bulk}). The non-zero components of the normals are 
$$^{(\tau)}n_{\tau}=\frac{1}{\sqrt{g^{\tau \tau}}}\,,\quad ^{(r)}n_{r}=\frac{1}{\sqrt{g^{rr}}}\,.$$ 
With these we calculate the two extrinsic curvatures $^{(\tau)}K_{\mu\nu}$ and $^{(r)}K_{\mu\nu}\,.$  Then we put $r= b\, x$ and $\tau=0$
as before. As the entangling surface is located at $r=0, \tau=0$  we further do an expansion around $x$ followed by an expansion in $n.$  Now $^{(\tau)}K_{a b}=0$ whereas $^{(r)}K_{ab}=A(x,n,\rho)$ is some function of $x$ , $n$ and $\rho\,.$  First we expand it around $x=0$ and then we do an expansion around $n=1\,.$  We find that  $^{(r)}K_{ab}=0\,.$  

Thus effectively the Fefferman-Graham construction is the same as the Iyer-Wald prescription, provided we take the limits in the manner prescribed in \cite{solo}. The replacement of $r K_{ij} dx^i dx^j$ by $r^n K_{ij} dx^i dx^j$ plays a key role in this construction. Recall that this was needed to keep the boundary Ricci scalar finite.  Also another important point to notice that for the squashed cone metric there is no time like killing vector as the metric  components are dependent on $\tau\,.$ The Wald-Iyer prescription calls for calculating the Wald functional in the context of  black hole  entropy where there exists a time like killing vector.  But  in the metric (\ref{metric1}) the  $cos(\tau)$ factor which breaks the time translational symmetry is accompanied by a factor of $r^{n}\,.$ In our calculation we have taken the $r\rightarrow 0$ limit first and then the $n\rightarrow 1$  limit. Thus the $cos(\tau)$ multiplied by $r^{n}$ is suppressed in this way of taking limits. For this reason we  have an approximate time-translational symmetry in our new space time.\par  Upto this point the discussion is independent of the choice of $b$. Now when one wants to evaluate the Wald entropy functional with this squashed cone  metric one needs to specify $b$ as mentioned in the previous sections for the sphere and the cylinder to obtain the correct universal terms.  As there is no integral over $r$ in the Wald entropy functional, the final result obtained  will be $b$ dependent as we have found and hence we have to choose $b$ accordingly.  

\subsection{Universality in Renyi entropy}
 In \cite{smolkin,galante, Perlmutter:2013gua, inprep} it was shown that for spherical entangling surfaces in four dimensions the Renyi entropy has a universal feature. Namely 
$$
\partial_n S_n|_{n=1}\propto c_T \,.
$$
In four dimensions $c_T \propto c$, the Weyl anomaly.
If we use eq.(\ref{ren}) and identify it as the expression for $S_n$ with the choice for $b$ given below it{\footnote{In order to get the proportionality constant to work out, we will need to adjust $\partial_n\beta(n)|_{n=1}$ in $b$.}, then we indeed find that this is true! This also works for the six and eight derivative examples. Thus this approach enables us to check some information away from $n=1$. Further, as a bonus, we can predict what happens in the case of a cylindrical entangling surface where holographic results for the Renyi entropy are not available. If we use eq.(\ref{ren2}) or its analog for the six and eight derivative examples, we find that $\partial_n S_n|_{n=1}\propto c_T$ still holds. It will be interesting to explicitly verify this in  field theory.

\section{Equation for the entangling surface}
In this section we turn to the derivation of the equation for the entangling surface following \cite{maldacena}.  Until now, we were interested in the leading order solution since this captured the universal term in EE. However, following the method proposed by LM, it is possible to derive the equation for the entangling surface which will carry information about how the surface extends into the bulk. The essential idea is to look at the singular components of the equations of motion arising due to the conical singularity and set them to zero. This was considered in \cite{ab, Chen} in the context of four derivative gravity.  We briefly review the necessary results below\footnote{Note that only the $rr$ component of the equations of motion was considered in \cite{Chen}.}.
 We start with the following metric,
\be \label{lin}
ds^{2}= e^{2\rho} (dr^{2}+r^{2}d\tau^{2})+(h_{ij}+ r \cos(\tau )\,{}^{(r)}\K_{ij}+r \sin(\tau)\,{} ^{(\tau)}\K_{ij})dx^{i}dx^{j}\,,\\
\ee
where, $\rho= -\epsilon \ln r$ and $n=1+\epsilon\,.$  The entangling surface is located at $r=0,\tau=0\,.$   We linearize  the equation of motion taking this metric $g_{\alpha\beta}$ and a fluctuation $\delta g_{\alpha\beta}$ of the type $\delta g(\tau)=\delta g(\tau+2\pi)\,.$ On general grounds we will get divergences of the type $\frac{\epsilon}{r}\,, (\frac{\epsilon}{r})^{2}\,.$  Setting these divergences to zero we get the minimal surface condition. Also following the periodicity argument in \cite{maldacena} we set the contribution coming from $\delta g$ which  is of the type $-\frac{1}{2}(1- 2 f_{\infty}\lambda)\Box g_{\alpha\beta}$ to zero. Below we list all the $\epsilon$-dependent divergences that arise in Gauss-Bonnet gravity. The equations of motion corresponding to the action (\ref{R}) with $\lambda_{1}=\lambda_{3}=\lambda\,\rm{and}\,\lambda_{2}=-4\lambda$  are given by,
\begin{align}
\begin{split} \label{eomfull}
G_{\alpha\beta}-\frac{6}{L^{2}}g_{\alpha\beta}-\frac{ L^{2}{\bf{\lambda }}}{2} H_{\alpha\beta}=T_{\alpha\beta}
\end{split}
\end{align}
where,$$
G_{\alpha\beta}= R_{\alpha\beta}-\frac{1}{2} g_{\alpha\beta}  R\,
$$
and
$$
 H_{\alpha\beta}=4  R_\alpha^\delta  R_{\beta \delta}-2  R  R_{\alpha\beta}  -4 R^{\delta \sigma} R_{\delta \alpha \beta \sigma} -2 R_{\alpha \sigma \delta \mu}{ R_\beta}{}^{\sigma \delta 
\mu}+\frac{1}{2} g_{\alpha\beta} GB\,.$$
\underline{Divergences in the $rr$  component}:
\begin{align}
\begin{split}
&-\frac{\epsilon }{r}\, \K-\frac{\lambda L^{2}\epsilon}{r}\big[\K\R-2 \K_{ij}\R^{ij}+r^{2\epsilon}(-\K^3+3 \K \K_2
-2\K_3)\big]\,.
\end{split}
\end{align}
\underline{Divergences in the $r\,i$  component }:
\begin{align}
\begin{split}
&-\frac{\lambda L^{2}\epsilon}{r}r^{2\epsilon}\big[2\K\nabla_j( \K^j_i)-2\K \nabla_i (\K) +2\K^j_i\nabla_j (\K)-2\K_{ij}\nabla_k (\K^{kj})+2\K_{kj}\nabla_i (\K^{kj})-2 \K_{jk}\nabla^j (\K^{k}_i)\big]\,.
\end{split}
\end{align}
\underline{Divergences in the $i\,j$ component} :
\begin{align}
\begin{split}
&4\lambda L^{2}\big[\frac{ \epsilon}{r}r^{4\epsilon}(\K_{ij}\K_2-2\K_{ik}\K^{kl}\K_{lj}+\K_{il}\K^{l}_{j}\K-\K\K_2h_{ij}+\K_3h_{ij})
+\frac{\epsilon^{2}}{r^2} r^{4\epsilon}(\K^2 h_{ij}-2\K\K_{ij}-\K_2h_{ij}+2\K_{ik}\K^{k}_{j})\big]\,.
\end{split}
\end{align}
$\R, \R_{ij}$ etc are made up of the metric $h_{ij}$\,, $\K_{2}=\K_{ab}\K^{ab}$ and $\K_{3}=\K_{ac}\K^{cb}\K^{a}_{b}$\,. Now to get the minimal surface condition we have to set all the divergences in the equation of motion to zero. The immediate question is how to handle the $r^{2\epsilon}$ terms which were absent in Einstein gravity considered in \cite{maldacena}.
Here we can proceed in two ways. Firstly, we can take the limit $\epsilon\rightarrow 0$ so that $r^{2\epsilon}\rightarrow 1$. This is what was implicitly done in \cite{ab, Chen}. Then we will be left over with divergences in all the components of the equations. In order to proceed, we could 
 assume the following as in \cite{ab} that $O(\K) \sim \alpha r/\epsilon$ where $\alpha\ll 1$, then the $ij$, $ir$ components go to zero. In that case $O(\K^3)\ll O(\K)$ so the $\K^3$ terms can be dropped. Thus finally we get,
\be \label{mincond}
 \K+\lambda L^{2}\big[\K\R-2 \K_{ij}\R^{ij}\big]=0\,.\\
\ee 
This matches with what follows from the Jacobson-Myers functional \cite{higher}. 
However, in order to do this consistently we needed to assume a weak extrinsic curvature limit. The other alternative is to consider the $r\rightarrow 0$ with $\epsilon\rightarrow 0$ limit in a way that we have a small parameter $r^{2\epsilon}$ in front of all the offending terms. In detail, if we demand $\epsilon/r \sim 1/\hat \epsilon, r^{2\epsilon}\sim \hat\epsilon^{1+\upsilon}$  with $\upsilon>0$, then taking the limit $\hat\epsilon\rightarrow 0$ and demanding that the equations are satisfied will lead to eq.(\ref{mincond}). 

We could alternatively have started with the following metric which is motivated by the regularization considered in \cite{solo},
\be\label{solo1}
ds^{2}=f(r,b) dr^{2}+r^{2} d\tau^{2}+  [h_{ij}+ r^{n} \cos(\tau )\,{}^{(r)}\K_{ij}+r^{n} \sin(\tau)\,{} ^{(\tau)}\K_{ij}]dx^{i}dx^{j}\,.\\
\ee
$f(r,b)$ is same as before and we have put in a factor of $r^{n}$ in front of the extrinsic curvature terms.  As explained before, all calculations with this metric need to be done by considering $r\rightarrow 0$ first and then $n\rightarrow 1$. Moreover\footnote{We thank an anonymous referee for pointing this out.}, this metric is related to the metric in eq.(\ref{lin}) around $r=0$ by a coordinate transformation $r\rightarrow r^n$ so would lead to the same results as above.


We will leave the analysis for the general four derivative theory, the six and eight derivative cases for future work {\footnote{In spite of computer help, this appears to be extremely tedious. For the six derivative case, the gravity equations can be found in \cite{Sinha:2010pm, quasi}. }. For the general four derivative theory, the method in \cite{maldacena} cannot be applied directly since it needs the contributions from metric fluctuations to vanish. In the $C_{abcd}C^{abcd}$ case, this does happen \cite{ab}. However, in this case the $O(r^2)$ contributions in the $g_{ij}$ metric become important \cite{ab}. These terms are regularization dependent--for example we could have replaced $r^2$ by $r^{2n}$ or left it as it is. Due to these complications we leave this interesting case for future work. 

\section{Discussion}
In this paper we showed the following:
\begin{itemize}
\item The newly proposed regularization in \cite{solo} yields the expected universal terms in the EE in higher derivative gravity theories dual to four dimensional CFTs. We considered the Fefferman-Graham metric with the regularized metrics in \cite{solo} as the boundary metric. Then we computed the generalized gravitational entropy as proposed in \cite{maldacena}. The universal log terms worked out to be as expected. We showed that upto the order we are interested in, the singularities in the metric are mild. As pointed out in \cite{maldacena} we could also have done a conformal transformation of the boundary metric with conical singularity such that it is non-singular and then done the calculation. We expect the results to be identical.

\item We computed the Wald entropy on the $r=0=\tau$ co-dimension 2 surface in the Fefferman-Graham metric and found that it gives the correct universal terms for both spherical and cylindrical surfaces. In order to get the expected results, we needed to choose a surface dependent but theory independent regularization parameter. Recall that in bulk AdS space, from the entropy functional way of computing EE in Lovelock theories, one needed to use the JM entropy functional which differed from the Wald entropy functional by extrinsic curvature terms. These extrinsic curvature terms are important to get the correct universal piece for any entangling surface with extrinsic curvature. However, the entropy functional for an arbitrary theory of gravity is not readily available. On the other hand, the observation that the Wald entropy in the squashed cone background as computed this paper leads to the expected universal terms opens the way to computing EE in an arbitrary higher curvature theory in even dimensions. Of course, in order to get the full entangling surface in the bulk, one still needs to first derive the relevant entropy functional and then minimize it.

\item We also showed that the entangling surface equations are the same as what comes from the JM entropy functional without the small extrinsic curvature condition needed in \cite{ab}. The essential point that enables this is to consider the $r\rightarrow 0$, $n\rightarrow 1$ limits in a way that lets $r^{n-1}\rightarrow 0$ rather than $r^{n-1}\rightarrow 1$ as was implicitly done in \cite{ab,Chen}. The considerations of the metric in eq.(\ref{solo1}) makes this somewhat clearer since all calculations in this metric need the limits to work this way.

\end{itemize}

There are several open problems. A justification for the choice of the surface dependent but theory independent regularization parameter in the calculation of Wald entropy has to be found.  In this paper we have  considered only spherical and cylindrical surfaces. But we expect that our method will work for any arbitrary surface.  It will be nice to determine a general form of $b$ for an arbitrary surface. We have extracted the logarithimic term from the Wald entropy as it requires only  information about the bulk space time around the boundary. 
 Although we have demonstrated that the regularized squashed cones of \cite{solo} can be used to compute EE, a naive application of this procedure would not work for Renyi entropies \cite{head, smolkin, fur1} for general $n$ although the starting metric is regular. Except near $n=1$, where we saw that the universality in Renyi entropy \cite{smolkin, galante, Perlmutter:2013gua, inprep} pertaining to $\partial_n S_n|_{n=1}$ bears out, the result for a general $n$ would be regularization dependent--for instance we will need to know details about $f(r,b)$ away from $r=0$. This problem may be interlinked with the previous one. In both cases, presumably global information of the metric is needed to fix the regularization ambiguities. Recall that in the calculation of the Renyi entropy for spherical entangling surface in \cite{smolkin} the periodicity of the time coordinate was fixed by knowing the relevant temperature of the hyperbolic black hole. In order to extract this information, it is necessary to know the  bulk geometry everywhere. In even dimensions the Fefferman-Graham expansion breaks down and hence a different approach may be needed to compute Renyi entropy. In odd dimensions, in principle it is possible to continue the expansion \cite{bhss} but in practice this appears very hard.

Whether EE can be thought of as a Noether charge needs further investigation. Our findings in this paper seems to suggest that this may indeed be true. The Fefferman-Graham metric is the analog of the Iyer-Wald metric used to compute the entropy for dynamical horizons. Our conjecture then is that the Wald entropy (after appropriately fixing the regularization) evaluated on the $r=0=\tau$ co-dimension two surface in the Fefferman-Graham metric is going to capture the expected universal terms for any entangling surface.

\section*{Acknowledgments} We thank Janet Hung, Apratim Kaviraj,  and especially Dmitri Fursaev,  Rob Myers and Sasha Patrushev for discussions and correspondence. We thank Rob Myers for sharing \cite{inprep} with us. AS is partially supported by a Ramanujan fellowship, Govt. of India.

\end{document}